\documentclass[11pt,final,journal,twoside]{IEEEtran}
\usepackage{blindtext}
\usepackage{titling}
\usepackage{array}
\preauthor{\begin{center}
    \large \lineskip .75em%
        \begin{tabular}[t]{>{\centering\arraybackslash}p{.45\textwidth}}}
\postauthor{\end{tabular}\par\end{center}}
\makeatletter
\renewcommand\and{
  \end{tabular}%
  \hfill
  \begin{tabular}[t]{>{\centering\arraybackslash}p{.45\textwidth}}}
  \makeatother
\usepackage{amsfonts,amsmath,amssymb,amsthm} 
\usepackage{latexsym} 
\usepackage{mathrsfs}
\usepackage{graphicx}
\usepackage{indentfirst}
\usepackage{fullpage}
\usepackage{url}
\usepackage{multicol}
\usepackage{color}
\usepackage{tikz} 
\usepackage{hyperref}

\newcommand{\F}{\mathbb{F}}

\newcommand{\N}{\mathbb{N}}

\newcommand{\HH}{\mathcal{H}}

\newcommand{\Y}{\mathcal{Y}}
\newcommand{\X}{\mathcal{X}}

\newtheorem{thm}{Theorem}

\theoremstyle{remark}

\newtheorem{ex}[thm]{Example}
\newtheorem{defi}[thm]{Definition}

\newcommand{\rmv}[1]{}

\title{Mathematical LoRE: Local Recovery of Erasures using Polynomials, Curves, Surfaces, and Liftings
\thanks{
Haymaker is with the Department of Mathematics and Statistics, Villanova University (email: \texttt{kathryn.haymaker@villanova.edu}). L\'opez and Matthews are with the Department of Mathematics, Virginia Tech (email: \texttt{\{hhlopez, gmatthews\}@vt.edu}). 
Malmskog is with the Department of Mathematics \& Computer Science, Colorado College (email: \texttt{bmalmskog@coloradocollege.edu}).  
Pi\~nero is with the Department of Mathematics, University of Puerto Rico - Ponce (email: \texttt{fernando.pinero1@upr.edu}).
The National Science Foundation partially supported the second (DMS-2201094 and DMS-2401558),  third  (DMS-2137661), and  fourth (DMS-2201075) authors. The fourth author is also partially supported by the Commonwealth Cyber Initiative. The team acknowledges support from the American Institute of Mathematics SQuaRES program. }}

\author{Kathryn Haymaker, Hiram H. L\'opez, Beth Malmskog, Gretchen L. Matthews,  and Fernando Pi\~nero}

\date{}

\begin{document}

\maketitle

\begin{abstract}
Employing underlying geometric and algebraic structures allows for constructing bespoke codes for local recovery of erasures. We survey techniques for enriching classical codes with additional machinery, such as using lines or curves in projective space for local recovery sets or products of curves to enhance the availability of data.   
\end{abstract}

\section{Introduction and Motivation} \label{section:intro}

Algebraic, geometric, and combinatorial structures have long supported the development of codes for error correction and erasure recovery in which these mathematical structures guarantee the desired properties of the code. Prime examples are  Reed-Solomon codes \cite{Reed_Solomon_60}, in which codewords are defined by evaluating polynomials of bounded degree at elements of a finite field. Reed-Solomon codes have seen widespread use since their introduction in 1960, from satellite communications to QR codes. However, the code length is bounded by the alphabet size, meaning no nontrivial binary Reed-Solomon codes exist. The desire to decouple the code length and alphabet size is natural, especially when combined with the search for long codes inspired by Shannon's Theorem. This aspiration led to the breakthrough construction of algebraic geometry codes in the 1970s by V. D. Goppa \cite{Gop82}. Algebraic geometry codes are built from curves or higher-dimensional varieties over finite fields, which can have more rational points than the field size, leading to codes that can be much longer than Reed-Solomon codes over the same field. The underlying vector spaces of functions gave rise to codes with exceptional parameters, including those exceeding the Gilbert-Varshamov bound \cite{TVZ}. Refinements of these structures can lead to codes amenable to the local recovery of erased data\rmv{, increasing the availability to supply the same data to multiple users simulatenously}. 

Traditionally, the decoder takes a received word as input and attempts to correct any errors or recover any erasures introduced during the transmission. This classical scenario assumes access to the entire received word. Distributed networks, cloud storage, and blockchains are formed from hundreds or thousands of servers. So, it is impractical to assume access to all coordinates of the received word in the decoding process. Hence, it is desirable to design codes that can recover a few codeword symbols by accessing only a few other symbols of the received word. Such a code is called a locally recoverable code, or LRC.

LRCs provide smart data storage across a network with limited recovery and repair traffic. A small recovery set reconstructs unavailable or lost symbols. The rich underlying mathematical structures that gave the world Reed-Solomon codes and their higher-genus relatives (algebraic geometry codes) aid in the design of LRCs. In Section \ref{section:tools}, we review some algebraic and geometric structures and how they  define LRCs. 

One might set various objectives while considering codes for local recovery. One goal is to have different (often many) disjoint recovery sets to ensure high data availability. Thus, even with multiple erasures, there is a recovery set with enough data to recover the missing symbol(s). The availability measures the largest number of disjoint recovery sets for all coordinates. We will see rich structures that give rise to codes with availability in Section \ref{section:availability}. 

Another objective is to minimize the size of the largest recovery set -  called the locality of the code - to reduce the network traffic involved in recovering a failed node or erasure. It is natural to ask if these codes are MDS or meet other classical bounds. 
Of course, the additional structure required on a classical code to achieve local recovery can adversely impact the code's overall efficiency and error-correcting capability. For this reason, while traditional bounds for error-correcting codes apply to locally recoverable codes, they are not tight. Instead, bounds for LRCs take into account locality. These bounds are discussed in Section \ref{section:bounds}. A conclusion may be found in Section \ref{section:conclusion}.

\section{Algebraic and Geometric Tools} \label{section:tools}

An approach to developing an LRC is to add an underlying algebraic or geometric structure to a classical code. These structures allow additional machinery, such as the trace function of an extension field to minimize data transmission, lines or curves in the projective space to create recovery sets, or products of curves to enhance availability. Polynomials, curves over finite fields, and other algebraic or algebraic geometric objects have a long history in coding theory. We will use standard coding theory notation throughout, working over the finite field $\F_q$ with $q$ elements. An $[n,k,d]$ code over the alphabet $\F_q$ is a $k$-dimensional $\F_q$-subspace of $\F_q^n$ such that any two different elements (called codewords) differ in at least $d$ coordinates. 

Consider a vector space $V$ of $\F_q$-valued functions $V$ and points $P_1, \dots, P_n$ on a geometric object so that $f(P_i) \in \F_q$ for all $i \in [n]:=\{1, \dots, n\}$ for all functions in $V$. We will consider evaluation codes 
$$C(D, V):= \left\{ \left( f(P_1), f(P_2), \dots, f(P_n) \right) : f \in V \right\}$$ where $D=(P_1, \dots, P_n)$; that is, $C(D,V)=ev\left( V \right)$ where 
$$  
\begin{array}{cccl}
ev: & V & \rightarrow & \F^n \\
&f& \mapsto & \left( f(P_1), f(P_2), \dots, f(P_n) \right). 
\end{array}
$$ 
We recognize this construction giving rise to important families of codes:
\begin{itemize}
    \item Taking positive integers $k \leq n \leq q$, $V=\F_q[x]_{<k}$, the set of polynomials with coefficients in $\F_q$ of degree less than $k$, and $\left\{ P_1, \dots, P_n \right\} \subseteq \F_q$, $C(D, V)$ is the $[n,k,n-k+1]$ Reed-Solomon code $RS(n,k)_q$ over $\F_q$. 
    \item 
    Setting $\left\{ P_1, \dots, P_n \right\} = \F_q^m$ and taking $V=\F_q[x_1, \dots, x_m]_{<k}$ to be the set of polynomials with coefficients in $\F_q$ of total degree less than $k$ gives a Reed-Muller code. 
\item 
Considering $\F_q$-rational points $P_1, \dots, P_n$ on a curve $X$ over $\F_q$ and a divisor $G$ with support not containing any $P_i$ allows one to define an algebraic geometry code by taking $V=\mathcal{L}(G)$. 
\end{itemize}
We will see that fine-tuning the design choices $D$ and $V$ allow local recovery of erased data. 

\begin{defi}
    A code $C \subseteq \F_q^n$ has locality $r$ if for each codeword coordinate $i \in [n]$, there exists a set $R_i \subseteq [n] \setminus \{ i \}$ of other coordinates such that for all $c \in C$, 
$c_i = \varphi(c\mid_{R_i})$
for some function $\varphi : \F_q^r \rightarrow \F_q$ and $\mid R_i \mid = r$. The set $R_i$ (resp., $R_i \cup \{ i \}$) is called a {recovery set} (resp., repair group) for $i$.
\end{defi}

Because any $[n,k]$ code has locality $k$, we are interested in codes with locality $r<k$. The following Singleton-type bound describes the relationship between the standard code parameters (length, dimension, and minimum distance) and the locality for a linear code \cite{Gopalan}; a version for nonlinear codes is given in \cite{Papailiopoulos_Dimakis}.

\begin{thm} \label{singleton}
 The parameters of an $[n,k,d]$ LRC $C$ with locality $r$ satisfy
\begin{equation} \label{eq:singleton}
    k+d \leq n + 1 - \left(\left \lceil \frac{k}{r} \right \rceil -1 \right).
\end{equation}
\end{thm}
If $C$ has parameters meeting Bound (\ref{eq:singleton}), we say that $C$ is an optimal LRC. It is  immediate that the Bound (\ref{singleton}) reduces to the Singleton bound when considering an $[n,k,d]$ code to have locality $k$. 

Next, we will meet some families of LRCs, one from polynomials and another from curves, some of which are optimal (see Section~\ref{section:bounds}).

\subsection{Locally recoverable codes from polynomials}
To construct a polynomial LRC, we first consider evaluation codes, similar to Reed-Solomon codes, that rely on a particular polynomial set known as \textit{good} polynomials. Fix integers $k, n, q, r$ with $k \leq n \leq q$ and $r+1 | n$. The latter condition may be dropped, but we keep it for the convenience.

\begin{defi} \label{def:good}
A polynomial $g(x) \in \F_q[x]$ is said to be good if $\deg g=r+1$ and there exists a partition $\mathcal P$ of $\F_q$, $$\F_q=A_1 \overset{\cdot}{\cup} \cdots \overset{\cdot}{\cup} A_{\frac{n}{r+1}},$$ with parts $A_l$ satisfying     $|A_l|=r+1$ for every $l \in [\frac{n}{r+1}]$ and $g(\alpha)=g(\beta)$ for all $\alpha,$ $\beta \in A_l$.
\end{defi}
The idea is that a good polynomial reduces to a constant on each part of $\mathcal P$, so each of these parts forms a repair group. Tamo and Barg used good polynomials in \cite{TB} to define optimal polynomial LRCs. Rather than evaluating all polynomials of bounded degree, as in the Reed-Solomon case, the functions that give rise to codewords in a polynomial LRC are elements of 
 $$\mathcal{L}(g) := \left< g(x)^j x^i : \begin{array}{l} 0 \leq j \leq \frac{k}{r}-1, \\ 0 \leq i \leq r-1 \end{array} \right>.$$
 The polynomial LRC is $C_k(g):=C(\F_q, \mathcal L(g))$, meaning
 $$C_k(g):=
\left\{ \left( ev(f) \right) : f  \in \mathcal{L}(g) \right\} \subseteq \F_q^{n}.$$
Notice that every  $f$ in $\mathcal{L}(g)$ can be written as 
$f(x) = \sum_{i=0}^{r-1} {\sum_{j=0}^{\frac{k}{r}-1} a_{ij} g(x)^j}x^i$ for some $a_{ij} \in \F_q$. This expression allows for local recovery of an erased coordinate $f(\alpha_i)$ as follows. Suppose $ \alpha_i \in A_l$. Since $g(x)$ is a good polynomial, $g(\alpha_j)=b$ for some $b \in \F_q$ for all $\alpha_j \in A_l$. Hence, the polynomial $f$, when restricted to $A_l$, is 
$$f(x)_{\mid_{A_l}} =
 \sum_{i=0}^{r-1} \sum_{j=0}^{\frac{k}{r}-1} a_{ij} g(x)^j x^i  =
 \sum_{i=0}^{r-1} \sum_{j=0}^{\frac{k}{r}-1} a_{ij} b^j x^i,$$ 
a polynomial of degree at most $r-1$. Then $f(x)_{\mid_{A_l}}$ can be determined using $r$ values via interpolation, so that  the value $f(\alpha_i)=f(\alpha_i)_{\mid_{A_l}}$ is found.
In \cite{TB}, it is shown that $C_k(g)$ is an $[n,k, \geq n-k-\frac{k}{r}+2]$ code, making it an optimal LRC. This construction depends on the existence of good polynomials. We will see an example and then discuss some sources of them. 

\begin{ex} \label{ex:TB}
Consider $\F_{13}^*= \left\{1, 2, \dots, 12 \right\}$, the nonzero elements of  $\F_{13}$. 
Take $g(x)=x^3$, so $r+1=3 \mid 12$. Notice that 
$$
\begin{array}{ccccccc}
g(1)&=&g(3)&=&g(9)&=&1,\\  g(2)&=&g(5)&=&g(6)&=&8,  \\g(4)&=&g(10)&=&g(12)&=&12, 
\\g(7)&=&g(8)&=&g(11)&=&5. \end{array}$$ 
Set $A_1= \left\{ 1, 3, 9 \right\}, A_2=\left\{ 2, 5, 6 \right\}, A_3=\left\{ 4, 10, 12 \right\},$ and $A_4=\left\{ 7,8,11 \right\}.$
Then $$\F_{13}^* = \left\{ 1, 3, 9 \right\} \cup \left\{ 2, 5, 6 \right\} \cup \left\{ 4, 10, 12 \right\} \cup \left\{ 7,8,11 \right\}.$$
 Here, $$
\begin{array}{lll}
\mathcal L(x^3) &=& \left< (x^3)^jx^i : 0 \leq j \leq 2, 0 \leq i \leq 1 \right> \\
& = & \left< 1, x, x^3, x^4, x^6, x^7 \right>.
\end{array}
$$
The  code $C_6(x^3)$ can recover any symbol $f(i)$ using only 2 symbols $f(j)$, where $i, j \in [12]$. For instance, suppose $$(1,3,1,4,?,1,1,10,1,3,11,7)$$ is received from sent $ev(f)$, meaning  $f(5)$ is erased and  $f(x) \in \mathcal L(g)$, as shown in Figure \ref{fig:TB}.
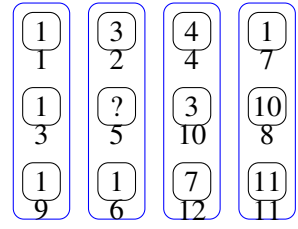
\begin{figure}
{{
\begin{center}
\begin{tikzpicture}[scale=0.5]
\foreach \i in {0,2,4,6}{
    \foreach \j in {0,2,4}{
        \draw[rounded corners] (\i,\j) rectangle (\i+1,\j+1);
    }
}
\foreach \k in {-0.25,1.75,3.75,5.75}
    \draw[blue,rounded corners] (\k,-0.5) rectangle (\k+1.5,5.25);
\path 
    \foreach \X [count=\Y] in {1,3,4,1,1, ?,3,10,1,1,7,11}
        {({Mod(\Y-1,4)*2+0.5},{-int((\Y-1)/4)*2+4.5})node{\X}};
\path 
    \foreach \X [count=\Y] in {1,2,4,7,3,5,10,8,9,6,12,11}
        {({Mod(\Y-1,4)*2+0.5},{-int((\Y-1)/4)*2+3.75})node{\X}};
\end{tikzpicture} \label{fig:TB}
\end{center}}}
\caption{Coordinates of a received word from polynomial-based LRC grouped according to the partition of $\F_{13}^*$ given by the polynomial $g(x)=x^3$.}
\end{figure}
To recover the symbol $f(5)$, we may use $f(2)=3$ and $f(6)=1$.  
The unique polynomial $h(x)$ of degree $r<2$ such that $h(2)=3$ and $h(6)=1$ is
$$h(x) = 3\left( \frac{x-6}{2-6}\right) + 1 \left( \frac{x-2}{6-2}\right) = 6x+4 \ \text{mod }13.$$ 
Then, the lost symbol is
$$f(5) = h(5) = 6(5)+4 = 34 = 8, $$
which is recovered using only two symbols rather than six required by a Reed-Solomon code of the same dimension.
\end{ex}

Example \ref{ex:TB} demonstrates how a good polynomial aids in local recovery, raising the question of how one finds good polynomials. 
Given a subgroup $H$ of $\F_q$, the polynomial $g(x):=\prod_{a \in H} \left(x-a\right)$ is a good polynomial with respect to the partition given by the cosets of $H$ \cite[Proposition 3.2]{TB}. In \cite{micheli}, Micheli demonstrates a Galois theoretical approach that explicitly constructs good polynomials. The key idea is that the number of partitions is linked to the number of totally split places of degree $1$ in $\F_q(x)/\F_q(t)$, where $t$ is transcendental and $x$ is a zero of $g-t$ in the algebraic closure of $\F_q(t)$. 

This idea of functions that reduce to low-degree polynomials is a theme throughout the algebraic constructions of LRCs. Next, we will see how curves over finite fields construct longer LRCs.

\subsection{Locally recoverable codes from curves}\label{LRCs from curves}
As described in the previous subsection, a major contribution of Tamo and Barg is that one can carefully create spaces of one-variable polynomials, including those of high-degree, so that each element of the evaluation set has a small helper set, on which these functions restrict to low-degree polynomials that can be interpolated using the helper set.  In 2015, Tamo, Barg, and Vladut  extended this idea to functions of more variables, evaluated them at selected points on a curve in a higher-dimensional space \cite{TBV_ISIT}; also see \cite{TBV}. The most basic version begins with a plane curve $\mathcal{X}$, defined by the vanishing of some polynomial $F\in\mathbb{F}_q[x,y]$. Concretely, points on $\mathcal{X}$ are pairs $(x,y)\in(\mathbb{F}_q)^{2}$ so that $F(x,y)=0$.  There may also be points at infinity, but for simplicity, we focus on the affine (non-infinite) points in this article.  Much like one could consider complex solutions to an equation with real coefficients, one may look for points on $\mathcal{X}$ in any larger field containing $\mathbb{F}_q$, and the set of points over a field $K/\mathbb{F}_q$ is specified by $\mathcal{X}(K)$.

To capture the intuition of the Tamo-Barg-Vladut construction, we begin with an illustrative example from \cite{TBV}.  Consider the Hermitian curve $\HH_q$ defined by the equation $x^q+x=y^{q+1}$.  We consider $x,y\in\F_{q^2}$. Note that $x^q+x$ is the field trace function that is a $q$-to-one map $\F_{q^2}\rightarrow\F_q$.  Also, $y^{q+1}$ is the field norm function that is a $(q+1)$-to-one map $\F_{q^2}^{\times}\rightarrow\F_q^\times$. For any $a\in \F_q^{\ast}:=\F_q \setminus \{ 0 \}$, there are $q$ $x$-values and $(q+1)$ $y$-values in $\F_{q^2}$ with $x^q+x=a=y^{q+1}$.  When $a=0$, we still have $q$ $x$-values but only a single $y$-value $(y=0)$ so that $x^q+x=0=y^{q+1}$. Counting the possibilities, there are $q^3$ solutions to this equation over the field $\F_{q^2}$.  For example, when $q=3$, we find 27 solutions to $x^3+x=y^4$, where $x,y\in\F_9$.  

The Hermitian curve has been well-studied by mathematicians and coding theorists. It is particularly interesting in coding theory because it is a \textit{maximal} curve, having as many points as mathematically possible for a curve of its complexity. Consider the set of functions on $\HH_q$ defined by 
\[V_l=\langle x^iy^j:0\leq i\leq q-2, 0\leq j\leq l\rangle,\]
where $l$ is a parameter that can be increased for a larger rate and decreased for a larger minimum distance in the final code. Let $S=\{P_1, P_2, \dots, P_{q^3}\}$ be the set of all affine points. Barg, Tamo, and Vladut showed that the evaluation code $C(S, V_l)$ is locally recoverable and determined its parameters.

To see that we have local recovery, consider the projection $\phi_y$ of points in $S$ to their $y$-coordinate.  Say $P_i=(\alpha,\beta)\in S$. Then $\phi_y^{-1}(\phi_y(P_i))$ is the set of all points on $\HH_q$ with $y$-coordinate $\beta$.  As above, there are $q$ different $x$-values over $\F_{q^2}$ so that $x^q+x=\beta^{q+1}$, so $|\phi_y^{-1}(\phi_y(P_i))|=q$.  We let $$R_{i}:=\phi_y^{-1}(\phi_y(P_i))\setminus\{P_i\}$$ be the recovery set for the $i$-th position in the code, corresponding to point $P_i$. Since all the members of $R_i$ have $y=\beta$, any $g\in V$ will act as a single variable polynomial $\tilde{g}=g(x,\beta)$ on $\phi_y^{-1}(\phi_y(P_i))$.  The degree of $\tilde{g}$ is at most $q-2$ by construction, so the value of $g(P_i)=\tilde{g}(\alpha)$ can be interpolated from the values of $g(P_j)\tilde{g}(x_j)$ for the $(q-1)$-elements $P_j$ of $R_i$. 

Continuing the example on $\HH_3$, we will choose $l=2$, yielding $$V=\langle 1, x, y, xy, y^2, xy^2\rangle,$$ a vector space of dimension 6.  For any $y$-value $\beta$ in $\F_9$, there are 3 $x$-values in $\F_9$ with $x^3+x=\beta^4$.  For each of the three points $(x,\beta)$, the recovery set for the position corresponding to one point is the positions of the other two points with $y=\beta$.  Thus, we have an LRC with $n=27$, $k=6$, and $r=2$.  
 
To state a more general construction, we introduce a little notation. One may define a field of rational functions on the points of $\mathcal{X}$ defined over $K$, denoted $K(\mathcal{X})=K(x,y)/(F)$. Each function in $K(\mathcal{X})$ has a representative $g\in K(x,y)$; two functions $g_1,g_2\in K(x,y)$ represent the same function in $K(\mathcal{X})$ if $g_1 = g_2 + hF$ for some $h\in K(x,y)$.  

Given two curves $\mathcal{X}_1$ and $\mathcal{X}_2$, a morphism $\phi:\mathcal{X}_1\rightarrow \mathcal{X}_2$ is a function defined by polynomials in $\mathbb{F}_q$ so that $\phi(\mathcal{X}_1(\mathbb{F}_q))=\mathcal{X}_2(\mathbb{F}_q)$. If such a map exists, we call $\phi$ a cover of $\mathcal{X}_2$ by $\mathcal{X}_1$.  Given a function $g\in K(\mathcal{X}_2)$, $g$ also defines a function on $\mathcal{X}_1$ by composition with $\phi$.  Thus there is a natural inclusion $K(\mathcal{X}_2)\subseteq K(\mathcal{X}_1)$. According to the primitive element theorem, we know that there is some $y\in K(\mathcal{X}_1)$ so that $K(\mathcal{X}_1)\equiv K(\mathcal{X}_2)(y)$, where there is some algebraic relation between $y$ and the elements of $K(\mathcal{X}_2)$.  Barg, Tamo, and Vladut's key observation was that we may create a locally recoverable code from any such covering of curves.

Let $d_{\phi}$ be the degree of the map $\phi$.  Let $S\subset \mathcal{X}_2(\mathbb{F}_q)$ such that, for each $P\in S$, \[|\phi^{-1}(P)\cap\mathcal{X}_1(\mathbb{F}_q)|=d_{\phi},\] i.e. each point in $S$ has a full fiber in the map $\phi$, with all points in the fiber defined over $\F_q$. We then define $T=\phi^{-1}(S)\subset \mathcal{X}_1(\mathbb{F}_q)$ to be the evaluation set for our code.  Let $B=\{g_1, g_2,\dots g_l\}$ be a basis for an $l$-dimensional linear space of functions in $\F_q(\mathcal{X}_2)$ having no poles in $S$.  Let \[V=\langle g_iy^j:, g_i\in B, 0\leq j\leq d_{\phi}-2\rangle,\]
an $\F_q$-vector space of functions defined on $T$.  
The evaluation code $C(T,V)$ is a locally recoverable code of dimension $k=l+d_{\phi}-1$ and locality $r=d_{\phi}-1$.  Local recovery is accomplished as follows.  Let $P_i\in T$, with $\phi(P_i)=Q\in S$. Let $$R_i=\phi^{-1}(\phi(P_i))\setminus \{P_i\}.$$We have that $g_j(P)=g_j(\phi(P))=g_j(Q)$ for all $P\in\phi^{-1}(\phi(P_i))$ and all $j\in[l]$.  Let $g\in V$. Imagine that position $i$ is erased from the corresponding codeword $g$.  We want to recover the $g(P_i)$ value using the recovery set $R_i$. We observe that $g|_{\phi^{-1}(Q)}$ is a polynomial in $y$ of degree at most $d_{\phi}-2$.  Thus $g(P_i)$ can be interpolated from the value of $g$ on the $d_{\phi}-1$ points in $R_i$.

\begin{ex}
  Let $$ \mathcal X_1:y^{11}+y=x^{12}$$ be the Hermitian curve and $$\mathcal X_2:y^{11}+y=x^3$$ its quotient
over $\F_{121}$ and the degree $4$ morphism
$$
\begin{array}{lclc}
\phi: & \mathcal X_1 & \rightarrow & \mathcal X_2 \\
&(x,y) & \mapsto & (x^{4}, y).
\end{array}
$$
Then
$C(G,g)$ is a 
$[1320, 660, 382]$ code with locality $3$. If we take instead $\mathcal X_3:y^{11}+y=x^4$, adjusting the morphism accordingly, we obtain a 
$[1320, 660, 277]$ with locality $2$. Using $\phi:\mathcal X_1 \rightarrow \mathbb P^1$, we obtain a 
$[1320, 660, 502]$ code with locality $11$. 
\end{ex}

This construction is remarkable in its flexibility, allowing LRCs with a wide range of localities to be constructed. Further, we see here an exciting application of algebraic-geometric structure and relationships to obtain desirable structures in codes.  The  Tamo-Barg polynomial LRCs described in the previous subsection are subcodes of Reed-Solomon codes, with length limited by field size.  By employing curves over finite fields, this construction allows for long evaluation-based LRCs to be constructed over smaller fields. 
In particular, we see that the length of codes defined over $\F_q$ from this construction can be up to $n \leq q+1+2g\sqrt{q}$ where $g$ is the genus of the curve $\mathcal X$. 

\section{Availability} \label{section:availability}
Locally recoverable codes are motivated by applications that are especially relevant in distributed storage, where servers will fail, and the information of a codeword symbol must be recovered from the remaining servers.  However, it may be the case that a particular codeword symbol is simply in high demand and therefore unavailable because of increased traffic.  Locally recoverable codes allow more users to access this symbol through its recovery set. If the symbol is in extremely high demand, it may be helpful to have additional disjoint recovery sets. This motivates the notion of availability in LRCs.

\begin{defi}    A locally recoverable code $C$ is said to have availability $t$ and locality $(r_1,r_2,\dots,r_t)$ if each codeword position has $t$ disjoint local recovery sets, with the $i$-th repair set for each position having a size at most $r_i$.  Such a code is called an LRC($t$).
\end{defi}

Locally recoverable codes with high availability are also exciting from a theoretical computer science perspective.  LRC($t$)s with $t=\Omega(n)$ are equivalent to locally decodable codes (LDCs) and have applications in private information retrieval and cryptography.

There are several approaches to constructing codes with availability by patching together LRCs.  For example, one can take a combinatorial approach, as in the product code arising from $t$ binary parity check codes suggested by Tamo and Barg in \cite{TamoBargBoundsAvailability}. Geometric constructions have again been extraordinarily useful in producing LRC($t$)s through natural geometric structures.  We discuss two major approaches here.

\subsection{Codes from fiber products}
We return now to the LRCs from covering maps of algebraic curves discussed in Section \ref{LRCs from curves}.  In this setting, locality arises from fibers in a covering map of curves.  To obtain availability, it would be useful to have a single curve with several covering maps to other curves, each of which would create a disjoint recovery set to each point.  In fact, an algebraic-geometric construction gives exactly such a curve, the fiber product.  Since these constructions and theorems are somewhat technical when stated in full generality, we omit the statements in favor of intuition and motivation. Intuitively, given $t$ curves $\Y_1,\dots, \Y_t$, each of which equipped with a covering map to another curve $$h_i:\Y_i\rightarrow\Y,$$\rmv{ as shown in Figure \ref{fig:fiber}} we can construct a curve $\X$ as the product of these $t$ curves, which will have the desired covering map $g_i: \X \rightarrow \Y_i$ such that $h_i \circ g_i=h_j \circ g_j$ for all $1 \leq i, j \leq t$.  We denote the fiber product by $$\X=\Y_1\times_{\Y}\Y_2 \times_{\Y} \dots \times_{\Y} \Y_t$$ and set $g:=h_i \circ g_i$. 
\rmv{
\begin{figure}[h] 
\centering
\begin{tikzpicture}[node distance = 2cm, auto] 
  \node (Y) {$\mathcal{Y}$};
  \node (Y1) [node distance=1.4cm, left of=Y, above of=Y] {$\mathcal{Y}_2$};
    \node (Ys) [node distance=1.4cm, right of=Y, above of=Y] {$\mathcal{Y}_t$};
   \node (X) [node distance=1.4cm, left of=Ys, above of=Ys] {$\mathcal{X}$};
 \node (Y2) [node distance=1.4cm, left of=Y1, left of=Ys, below of=X] {$\mathcal{Y}_1$};
  \draw[->] (Y1) to node {$h_2$} (Y);
   \draw[->] (Y2) to node [swap] {$h_1$} (Y);
  \draw[->] (Ys) to node {$h_t$} (Y);
   \draw[->] (X) to node [swap] {$g_1$} (Y2);
  \draw[->] (X) to node {$g_2$} (Y1);
  \draw[->] (X) to node {$g_t$} (Ys);
  \path (Y1) -- node[auto=false]{\ \ \ldots } (Ys);
  \draw[->, bend left] (X) to node {$g$} (Y);
\end{tikzpicture}
  \caption{The fiber product $\mathcal{X}$ of $t$ curves $\mathcal{Y}_j$. }
  \label{fig:fiber}
  \end{figure}}
The fiber product of curves is a curve that can be considered a generalized intersection of hypersurfaces defined by the equations for the curves $\Y_i$ in a higher-dimensional space.  Fiber products can be viewed abstractly from a category or scheme-theoretic perspective.  However, there is no need for advanced machinery to work with these objects.  The points of $\X(\F_q)$ are just tuples of points $(P_1,P_2,\dots, P_t)$ in the Cartesian product $$\Y_1(\F_q)\times \Y_2(\F_q)\times \dots \times \Y_t(\F_q)$$ such that $$h_1(P_1)=h_2(P_2)=\dots = h_t(P_t)\in\Y(\F_q).$$The fiber product can create a ``custom" curve, which is equipped with maps arising from the projections $h_i$.  It is one natural extension of the Barg, Tamo, Vladut construction of LRCs from covering maps of curves to LRC($t$)s.  In fact, these authors described the idea, gave parameter bounds, and constructed examples of LRC($2$)s from fiber products in \cite{TBV}.  Haymaker, Malmskog, and Matthews extended the construction to LRC($t$)s in \cite{fiber}, giving a slightly improved minimum distance.  

Here, we consider some instances of well-known fiber products that yield codes with availability. The first one provides availability two, and the second yields an arbitrary availability. Both are from families of curves with ``built-in'' recovery sets satisfying the disjoint repair property arising from factors in the fiber product. 

\subsubsection{Generalized Giulietti-Korchmaros (GK) curves}
 Consider the generalized Giulietti-Korchmaros curve $\mathcal X_N$ over $\F_{q^{2N}}$ for an integer $N\geq3$, which is the fiber product over $\mathbb{P}^1$ of two curves:
 \[ \mathcal{Y}_N: y^{q^2} - y =z^{\frac{q^N+1}{q+1}}. \]   
\[ \mathcal{H}_q: x^q+x=y^{q+1}. \]  
The number of $\mathbb{F}_{q^{2N}}$-rational points on $\mathcal{X}_N$ is 
\[ \#\mathcal{X}_N(\mathbb{F}_{q^{2N}} )= q^{2N+2} - q^{N+3} + q^{N+2} + 1. \] 
We can take $h_1:\mathcal{Y}_N\rightarrow \mathbb{P}^1_y$ to be the natural degree $\frac{q^N+1}{q+1}$ projection map onto the $y$ coordinate for affine points, with $\infty_{\mathcal{Y}_N}\mapsto \infty_{y}$  and $h_2:\mathcal{H}_q\rightarrow \mathbb{P}^1_y$ be the natural degree $q$ projection map onto the $y$ coordinate for affine points, with $\infty_{\mathcal{H}_q}\mapsto \infty_{y}$.
One may verify that this design gives rise to two disjoint recovery sets for each coordinate: one of size $q-1$ and another with $\frac{q^N+1}{q+1}-1$ elements. Even though the projections $g_1$ and $g_2$ hark back to the constructions using covers, since they arise from the fiber product, the recovery sets produced are disjoint. That would not typically occur if one begins with a curve $\mathcal X$ and projects onto different curves $\mathcal Y_1$ and $\mathcal Y_2$. 

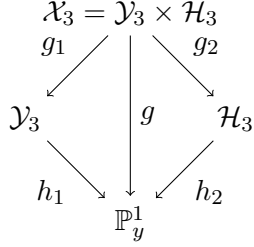
\begin{figure} 
\centering
\begin{tikzpicture}[node distance = 2cm, auto]
  \node (Y) {$\mathbb{P}^1_y$};
  \node (Y1) [node distance=1.4cm, left of=Y, above of=Y] {$\mathcal{Y}_3$};
   \node (Y2) [node distance=1.4cm, right of=Y, above of=Y] {$\mathcal{H}_{3}$};
   \node (X) [node distance=1.4cm, left of=Y2, above of=Y2] {$\mathcal X_3 = \mathcal{Y}_3\times\mathcal{H}_{3}$};
  \draw[->] (Y1) to node [swap] {$h_1$} (Y);
  \draw[->] (Y2) to node {$h_2$} (Y);
  \draw[->] (X) to node [swap] {$g_1$} (Y1);
  \draw[->] (X) to node {$g_2$} (Y2);
  \draw[->] (X) to node {$g$} (Y);
 \end{tikzpicture}
   \caption{Generalized GK curve over $\F_{729}$ as a fiber product, which gives rise to LRCs with availability $2$.}
    \label{fig:GK}
\end{figure}

\begin{ex} \label{E:GK_ex}
Consider the generalized Giulietti-Korchmaros curve $\mathcal X_3$ over $\F_{729}$ as pictured in Figure \ref{fig:GK}. Here, $N=3$ and $q=3$, and $\mathcal X_3$ is  
the fiber product over $\mathbb{P}^1$ of two curves:
\[ \mathcal{Y}_3: y^{9} - y =z^{7} \]  
\[ \mathcal{H}_3: x^3+x=y^{4}. \] 
There number of $\mathbb{F}_{729}$-rational points on $\mathcal{X}_3$ is 
\[ \#\mathcal{X}_3\mathbb{F}_{729}= 3^{8} - 3^{6} + 3^{5} + 1 = 6076, \]of which $6076-27=6049$ are evaluation points for the code.
The recovery sets are based on maps $h_1:\mathcal{Y}_3\rightarrow \mathbb{P}^1_y$  and $h_2:\mathcal{H}_3\rightarrow \mathbb{P}^1_y$,  the natural degree $7$ and degree $3$ projection maps mapping infinite points on $\mathcal Y_3$ and $\mathcal H_3$ to $\infty_{y}$, the point at infinity on $\Bbb P_y^1$. This setting gives rise to codes of length $6048$ in which each coordinate has two disjoint recovery sets: one of cardinality $6$ and the other of cardinality $2$. The dimension can be tailored to any value 
$$k=\left(\frac{q^N+1}{q+1}-1\right)(q-1)(l+1)=12(l+1)$$ by choosing a positive integer $l<q^{N+2}+q^{N+1}-q-1=320$, in which case we take 
 $$V= \left< x^i z^j y^{\kappa}:
 \begin{array}{l} 0 \leq i \leq 5,  0\leq j \leq 1,\\ 0 \leq \kappa \leq l, i,j, \kappa \in \mathbb{Z} \end{array} \right>.$$ For instance, setting $l=260$ gives a $[6048, 3132]$ code over $\F_{729}$, which has localities $2$ and $6$.
 \begin{figure} 
\centering
\includegraphics[scale=.15]{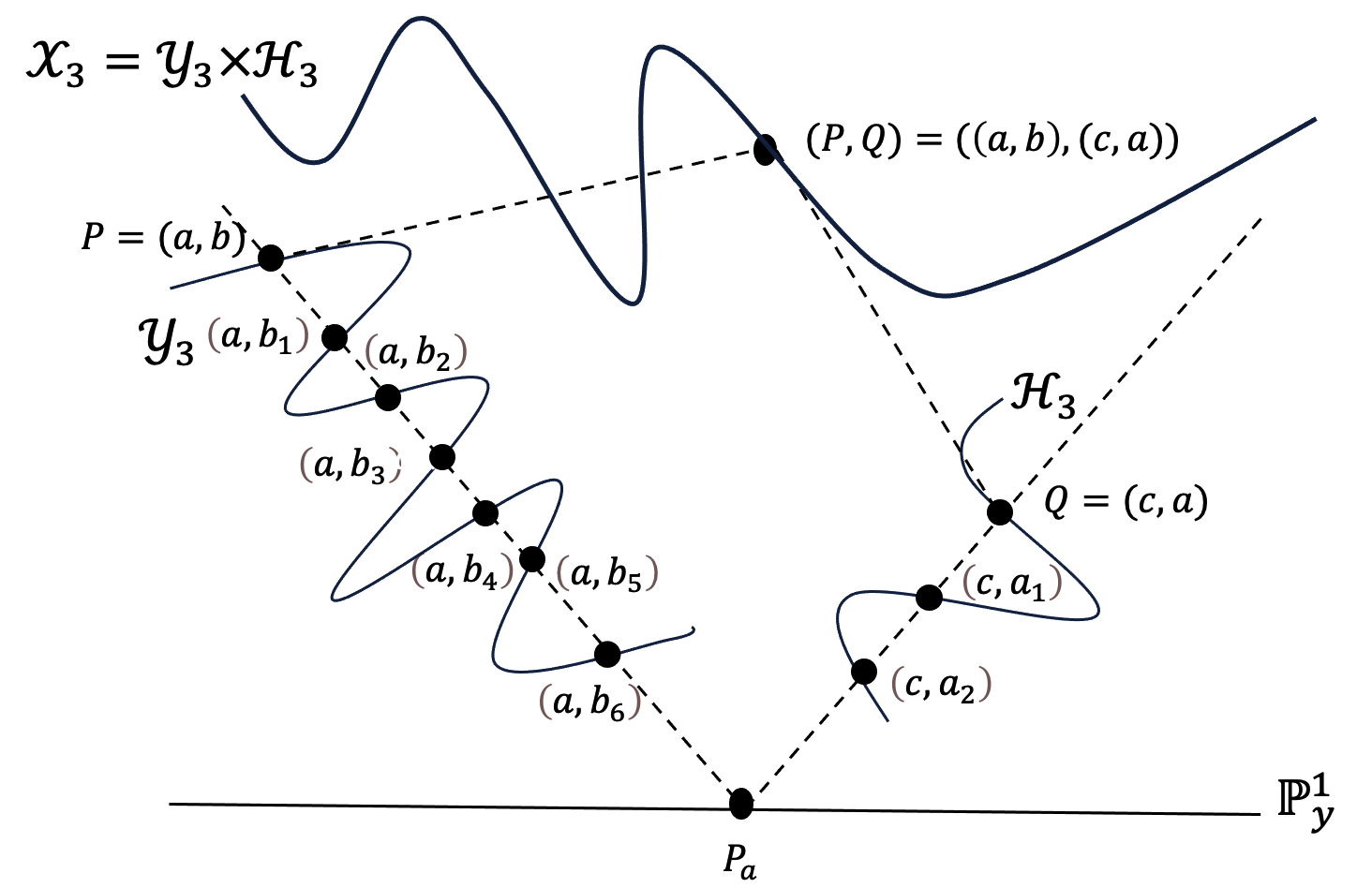}
   \caption{On the generalized GK curve $X_3$ over $\F_{729}$, two disjoint recovery sets for an erasure at the point $(P,Q)$ naturally arise from the fiber product structure, one comprised of points $(P,Q')$, where $Q' \neq Q$, and other other of points $(P',Q)$, where $P' \neq P$ as specified in Example \ref{E:GK_ex}.}
    \label{fig:GK2}
\end{figure}
 In particular, consider the point $(P,Q)$ on $\mathcal X_3$ where $P=(a,b)$ is a point on $\mathcal Y_3$ and $Q=(c,a)$ is a point on $\mathcal H_3$ as shown in Figure \ref{fig:GK2}. Then an erasure at $(P,Q)$ can be recovered using either the coordinates indexed by the set
 $$R_1:=\left\{ \begin{array}{l}
  ((a,b_1 ),Q),((a,b_2 ),Q),((a,b_3 ),Q),\\((a,b_4 ),Q),((a,b_5 ),Q),((a,b_6 ),Q)) \end{array} \right\}$$ or by the set $$R_2:=\left\{(P,(c,a_1 )),(P,(c,a_2 ))\right\}.$$
\end{ex}

As in this example, the generalized Giulietti-Korchmaros curves naturally yield availability $2$, meaning locally recoverable codes with two disjoint recovery sets for each symbol, one from each factor in the fiber product. \rmv{Given an erasure in a coordinate corresponding to a point $P$, it would be natural to first try to recover using $q-1$ symbols and if any are not available use the set of $\frac{q^N+1}{q+1}-1$ elements.}

\subsubsection{Artin-Schreier curves}

To design for a particular availability $t$, one may use a family of maximal curves which are 
fiber products of $t$ Artin-Schreier curves (themselves maximal curves) as studied by van der Geer and van der Vlugt \cite{vanderGeer1996}. 

\subsection{Lifted codes}
Polynomial LRCs and LRCs from curves have a key idea in common: restriction or projection naturally gives rise to recovery sets. Liftings, introduced by Guo, Kopparty, and Sudan \cite{guo_lifting}, suggest another approach using curves to design LRCs with availability. The idea is to choose a short code (for example, a Reed-Solomon code) and define a longer code by requiring that all codewords in the longer code restrict to codewords of the short code (on particular subsets). One version of this extends a Reed-Muller code with multivariate polynomials of total degree at most $q-2$ over $\mathbb{F}_q$ by adding all polynomials which restrict to single variate polynomials of degree at most $q-2$ on every line. The surprising insight of Guo, Kopparty, and Sudan is that adding these polynomials dramatically increases the code rate, yielding LRCs with high availability and good rates. This idea has been adapted to the Hermitian curves to yield Hermitian-lifted codes defined in \cite{lifted}. With Hermitian-lifted codes, intersections of lines and curves provide a high rate and high availability via algebraic-geometric constructions, which we will describe now. 
The choice of lines rather than quadratics or higher degree curves is intentional, motivated by the need for disjoint recovery sets. Because any two lines $L$ and $L'$ intersect in at most a single point $P$, choosing points from $L \setminus \{P \}$ and $L' \setminus \{P \}$ as recovery sets allows them to be naturally disjoint. 

We return to the Hermitian curve to describe the construction, though other curves have been explored similarly.  To build high-rate codes, codewords are defined using a carefully curated, as-large-as-possible set of rational functions from among the rational functions $\F_{q^2}(\mathcal H_q)$ on the Hermitian curve $\mathcal H_q$, in particular, those that restrict to low-degree polynomials on the intersections of the lines and the curve $\mathcal X$. Let 
 $L_{\alpha,\beta}(t):=\alpha t + \beta$ 
  with $\alpha,\beta \in \F_{q}$, with $\alpha \neq 0$. 
  
  \begin{ex}
      Consider the Hermitian curve $\mathcal H_4: y^4+y=x^5$ over $\F_{16}$ and the line $L:=L_{\alpha,\beta}(t)$
  with nonzero $\alpha,\beta \in \F_{16}$. Then the rational function given by the monomial $x^8y^2$ restricted to the line $L$ is 
$$
\begin{array}{lcl}
x^8y^2 \mid_L &=& t^8(\alpha t + \beta)^2\\
& = & \alpha^2 t^{10} + \beta^2 t^8
\end{array}
$$
according to the Freshman's Dream. While at first glance, it appears that $\deg \left( x^8y^2 \mid_L \right) =10$, further reduction is possible using the equation of the curve and the fact that 
$$
(\alpha t + \beta)^4+(\alpha t + \beta)=t^5
$$
for all points on both the line $L$ and the curve $\mathcal H_q$. In fact, in \cite{lifted}, it is shown that $\deg \left( x^8y^2 \mid_L \right) \leq 3$. Functions that reduce in such a way are the inspiration for Hermitian-lifted codes. 
  \end{ex}

  The set of all non-horizontal lines is $$\mathbb L:= \left\{ L_{\alpha,\beta}: \alpha, \beta \in \F_{q}, \alpha \neq 0 \right\}.$$ To consider which functions restrict to low-degree polynomials on the line  $L_{\alpha,\beta}$, we consider functions $f$
 modulo $$m_{\alpha, \beta, q}(t):= \left( \alpha t + \beta \right)^q+\left( \alpha t + \beta \right)-t^{q+1}$$ and $f \equiv h$ to mean $f \equiv h \mod m_{\alpha,\beta}$. 
 
\begin{defi}
The Hermitian-lifted code is $C(D, \mathcal F)$, where 
$$
\mathcal{F}:=  \left\lbrace f \in \F_{q^2}\left(\mathcal H_q \right) : \begin{array}{l} \exists g \in \F_{q}[t]_{\leq q-1}  \text{ s.t. } \\ f \circ L \equiv g \text{ }\forall L \in \mathbb{L} \end{array} \right\rbrace,
$$
$D= P_1+ \dots+ P_n $ is the sum  of $\F_{q^2}$-rational points on $\mathcal H_q$ other than $P_{\infty}$, and  $n=q^3$. Elements of $\mathcal F$ are called good for $\mathcal F$.
\end{defi}

Due to the underlying geometry of the Hermitian curve, any non-horizontal line $L$ through $P_i$ intersects $\mathcal{X}$ in  $q$ other $\F_{q^2}$-points. Each collection of $q$ such points acts as a recovery set for the coordinate associated with $P_i$. Moreover, there are $q^2-1$ such lines, one for each $\alpha \in \F_{q^2} \setminus \{ 0 \}$. It is then easy to see that the Hermitian-lifted code over $\F_{q^2}$ has length $q^3$, 
locality $q$, and availability $q^2-1$. 

Analyzing these lifted codes depends largely on understanding $\mathcal F$, the set of functions that restrict to polynomials of degree at most $q-1$ on the curve. Recall the set of functions on $\mathcal H_q$ with no poles other than $P_{\infty}$ is spanned by 
$$\bigcup_{m \in \N}  \left< x^i y^j : 0 \leq i \leq q, iq+j(q+1) \leq m\right>.$$
Certainly,  $x^ay^b \in \mathcal F$ provided $a+b < q$, but it turns out that there are many more functions. To get a handle on which rational functions are elements of $\mathcal F$, for a polynomial $f(t) \in \F_{q^2}[t]$, define
$$\bar{f}_{\alpha, \beta}(t) := f(t) \text{ mod } m_{\alpha, \beta}(t),$$
 the remainder resulting upon division of $f(t)$ by $m_{\alpha, \beta}(t)$, and
set
$\deg_{\alpha, \beta}(f) :=\deg(\bar{f}_{\alpha, \beta}(t))$.
Then
\[
\deg \left( \bar{f}_{\alpha, \beta}(t) \right) \leq q
\]
for all $f \in \F_{q^2}[t]$. 
We say that a monomial $x^ay^b$ is good for $\mathcal F$ if for all lines $L_{\alpha, \beta} \in \mathbb{L}$ with $\alpha \neq 0$,
$$\deg_{\alpha, \beta}(M_{a,b} \circ L_{\alpha, \beta}) \leq q-1.$$ 
Certainly,  monomials $x^a y^b$ with $a + b \leq q-1$ are good. In addition, some monomials $x^a y^b$ with $a + b \geq q$ are also good, namely those that happen to reduce to those of degree less than the locality on all lines; they are called sporadic. 

\begin{figure}[h!]
\centering
  \includegraphics[width=.8\linewidth]{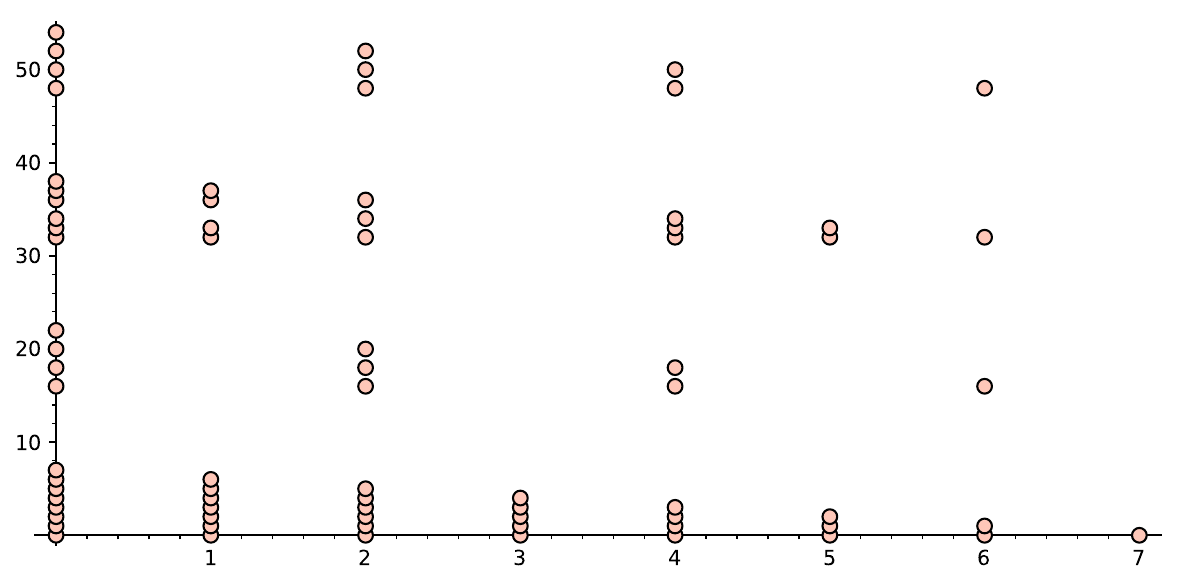}
  \caption{Exponent pairs $(a,b)$ with $x^ay^b$ for $q=8$ ($a$ is on horizontal axis) \cite{lifted} }
  \label{8fig}
\end{figure}

\begin{ex} \label{ex:Herm_lifted_8}
Consider the Hermitian curve $x^8+x=y^9$ over $\F_{64}$, which has 512 rational points other than the point at infinity, giving a code of length $n=512$. The monomials $x^ay^b$ with $a+b \leq 7$ are good for $\mathcal F$. One may also calculate sporadic good monomials using \cite{sagemath}. The exponent vectors $(a,b)$ satisfying this condition form a triangle as shown in Figure \ref{8fig}. 
\end{ex}

While computational tools such as \cite{sagemath} may be used to determine the sporadic monomials for particular values of $q$, combinatorial methods are needed for general $q$. Applying  Lucas's Theorem as in \cite[Theorem 10]{lifted}, we see that $x^ay^b$ with $a \leq q-1$, $b \leq q^2-1$, $a+b\geq q$ such that there exists $l, i \in [l]$, $0 \leq s \leq i-1$  so that 
$b=wq+b'$ with $b'< 2^{l-1}$, $2^i \mid w,  a < 2^{l-1}$, and no $2^s$ term in binary expansions of $a$ and $b'$ is a sporadic monomials. This observation allows one to prove the following.

\begin{thm} \label{thm:Herm_lifted}
Suppose that $q \geq 4$ is a power of $2$.  Then the Hermitian lifted code $C(D,\mathcal F)$ is a code of length $q^3$, locality $q$, availability $q^2-1$, and rate of at least $0.007$.
\end{thm}

\begin{ex}
Continuing Example \ref{ex:Herm_lifted_8}, we see that Theorem \ref{thm:Herm_lifted} indicates that the code $C(D, \mathcal  F)$ has dimension 75, with basis the set of monomials $x^ay^b$  plotted in Figure \ref{8fig}. In contrast, the comparable non-lifted one-point Hermitian code $C_{8,63}$ has dimension 36, making the rate of the Hermitian-lifted code at least $C$ $\frac{75}{512} \approx 0.15$ whereas the rate of the comparable one-point code is $\frac{36}{512}\approx 0.07$.
\end{ex}

Generalizing the Hermitian-lifted code construction to other curves can be a worthwhile but delicate process. We consider binary norm-trace lifted codes, as in \cite{nt_lifted_binary}, to see that. The norm-trace curve over $\F_{q^r}$ is 
$$
\mathcal X_{q,r}:   y^{q^{r-1}}+ y^{q^{r-2}}+ \dots +y^q + y = x^{\frac{q^r-1}{q-1}}; 
$$
meaning the norm of $x$ is the trace of $y$ where both the norm and the trace are taken relative to the extension $\F_{q^r}/\F_q$. Taking $r=2$ gives the  Hermitian curve $H_q: y^q+y=x^{q+1}$ over $\F_{q^2}$. The norm-trace curve over $\F_{q^r}$ has $q^{2r-1}$ rational points other than the point at infinity.

To mimic the Hermitian-lifted code construction for the norm-trace curve, one must determine the cardinalities of the intersections of lines with points on the curve. Restricting to the case $q=2$, we see
\begin{equation} \label{eq:binary_intersection_numbers}
\mid \left( L \cap \mathcal{X}_{2,r}\right)(\F_{q}) \mid = 2^{r-1} \pm 1
\end{equation}
for all lines $L$ \cite[Lemma 1]{nt_lifted_binary}. Hence, any non-horizontal lines $L$ through an evaluation point $P_i$ intersect $\mathcal{X}_{2,r}$ in at least $2^{r-1}-2$ other $\F_{2^r}$-rational points. This allows for locality $2^{r-1}-2$ and $2^r-1$ recovery sets for each coordinate. To define the binary-norm-trace lifted code, we need the appropriate set of functions, described next.

\begin{defi}
The binary norm-trace-lifted code is $C(D, \mathcal F_b)$ where 
$$
\mathcal{F}_b:=  \left\lbrace f \in \F_{2^r}\left(\mathcal X_{2,r} \right) : \begin{array}{l} \exists g \in \F_{2^r}[t]_{\leq 2^{r-1}-2}, \\ f \circ L \equiv g \text{ }\forall L \in \mathbb{L} \end{array} \right\rbrace,
$$
$D= P_1+ \dots+ P_n $ is the sum  of $\F_{2^r}$-rational points on $\mathcal X_{2,r}$ other than $P_{\infty}$, and  $n=2^{2r-1}$. 
\end{defi}

\begin{thm} \label{thm:nt_lifted}
Suppose that $q \geq 4$ is a power of $2$.  Then the binary norm-trace-lifted code $C(D,\mathcal F_b)$ is a code of length $2^{2r-1}$, locality $2^{r-1}-2$, availability $2^r-1$ with an asymptotic rate of $0.25$.
\end{thm}

 \rmv{Table \ref{table:Herm_norm_trace_q8} captures relevant parameters of Hermitian-lifted and binary norm-trace-lifted codes $\F_{64}$, meaning comparing lifted codes constructed using $\mathcal H_8$ and $\mathcal X_{2,6}$. }

\begin{table}[!htbp]
\centering
\label{specific-table-2}
\begin{tabular}{|l|c|c|c|c|}
    \hline
    ($r=6$) & Norm-trace code & HLC & NTLC \\
    \hline
    Field size &  $64$ & $64$ & $64$ \\
    Locality &  $30$ & $8$ & $30$ \\
    Availability &  $63$ & $63$ & $63$ \\
    Length &  $2048$ & $512$ & $2048$ \\
    Dimension &  $240$ & $75$ & $465$ \\
    \hline
\end{tabular} \\ \ \\
\caption{Hermitian-lifted and norm-trace-lifted codes over $\F_{64}$}
\label{table:Herm_norm_trace_q8}
\end{table}

\begin{figure}[!htbp]
\centering
\includegraphics[scale=0.4]{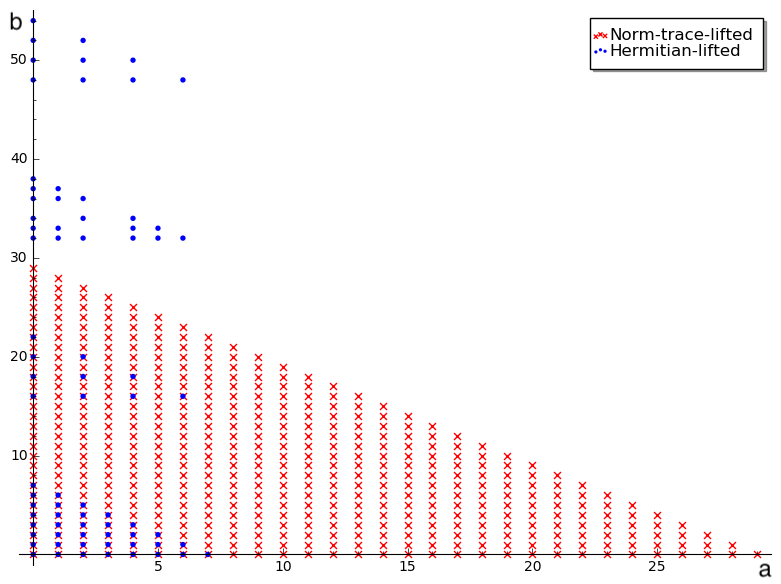}
\caption{Exponents of some good monomials $x^ay^b$ for Hermitian-lifted and norm-trace-lifted codes over $\F_{64}$ defined using $\mathcal H_8$ and $\mathcal X_{2,6}$ \cite{nt_lifted_binary}}
\label{fig:compare_HLC-NTLC_q8}
\end{figure}

It is natural to consider comparisons between the lifted constructions using the Hermitian and norm-trace codes.
Table \ref{table:Herm_norm_trace_q8} highlights the larger relative locality of the norm-trace-lifted codes, which might not be desirable, given that a larger percentage of coordinates are involved in local recovery than in the comparable Hermitian case. Even so, there is an upside. The greater locality means a more controlled set of good monomials, as seen in Figure \ref{fig:compare_HLC-NTLC_q8}.

\hspace{0mm}

\section{Bounds and Some Optimal Constructions} \label{section:bounds}
The classical MDS bound states that for an $[n,k,d]$ code, $d \leq n-k+1$. When locality $r$  is taken into account, the bound was refined to $d \leq n-k+1  - \left( \lceil \frac{k}{r} \rceil -1\right)$ as in Theorem \ref{singleton}. 

Note that the polynomial LRCs with alphabet $\F_q$ have length at most $q$. In contrast, longer LRCs from curves are possible but not optimal. It is natural to wonder if optimal LRCs are short. This question is reminiscent of the MDS conjecture for linear codes, prompting an interesting line of work resulting in optimal LRCs of lengths $n=q+1$ \cite{jin_ma_xing} using elliptic curves and $n=q+2\sqrt{q}$ \cite{li_ma_xing} using automorphism groups. 

When equality is attained in codes without locality,  the length is bounded by the alphabet size $n \leq q+1$ ($n \leq q+2$ for some cases when  $q$ is even). However, when considering optimal codes with locality, the length can be much larger than the alphabet size. For example, Luo, Xing, and Yuan proved in \cite{LYX18} that optimal LRCs of any length $n$ for $d=3,4$ exist. Moreover, in \cite{GXY}, Guruswami, Xing, and Yuan showed that if $d \geq 5$, optimal LRCs have length at most $O(dq^3)$. If $d = 5$, then optimal LRCs have length at most $O(q^2)$. 

For LRCs with availability $t$, an analog to the Singleton bound is: 
\begin{equation} \label{eq:availability-bound}
     d\leq n-k-\left\lceil\frac{t(k-1)+1}{t(r-1)+1}\right\rceil+2,
\end{equation}
which was introduced in \cite{wang2014repair} for linear codes and proved in \cite{rawat2016locality} for linear and nonlinear codes in which every information symbol has $t$ disjoint repair groups. In \cite{wang2014repair}, Wang and Zhang construct optimal linear LRC($t$)s for $n\geq k(rt+1)$. In general, it is an open question whether Bound (\ref{eq:availability-bound}) is tight, and the project of constructing optimal LRC($t$)s is ongoing. In \cite{tamo2016bounds}, the authors proved the following for LRC($t$)s in which every codeword symbol has $t$ disjoint repair groups: 
\begin{equation}\label{eq:avail-all}
    d\leq n-\sum_{i=0}^t \left\lfloor\frac{k-1}{r^i}\right\rfloor. 
\end{equation}

Bound (\ref{eq:avail-all}) is asymptotically better than Bound (\ref{eq:availability-bound}) but applies to a more restricted family of all-symbol locality codes. 



\subsection{Locally recoverable codes from surfaces: demonstrating that locally recoverable codes do not satisfy an MDS-like conjecture}
Defining LRCs from surfaces is a higher-dimension analog to the BTV construction using fibers of a map of curves. One approach is to define a morphism $\mathcal{X}$ from a surface $X$ in $\mathbb{P}^3$ to $\mathbb{P}^2$ using a projection map \cite{BHHMV}. Taking the fibers of $\mathcal{X}$ to be the helper sets, the construction results in a code of locality $r$. Codes can also be constructed by evaluating functions on a curve lying on a surface and using the geometry of the ambient surface to understand the code parameters \cite{SVV}. In fact, the authors of \cite{SVV} recast the Barg-Tamo-Vladut codes in light of this geometric viewpoint. 

Before delving into details, we discuss the significance of the initial constructions of codes from surfaces in \cite{BHHMV} in the evolution of understanding optimal LRCs.  Classical codes that meet the Singleton bound are maximum distance separable (MDS) codes. The ``trivial'' constructions of MDS codes of any length $n$ have distance $d=1, 2$, or $n$. All known nontrivial MDS codes have length $n<q+1$, except for even $q$ and $k=3$ or $k=q-1$ when there are codes with $n=q+2$. The MDS conjecture states that longer MDS codes over $\mathbb{F}_q$ do not exist. The conjecture is true for $q$ prime \cite{ball2012large}. However, the full MDS conjecture remains open. When attempting to form an MDS-type conjecture for LRCs, the examples in \cite{BHHMV} showed that the bound of $q+1$ would not directly translate. 

Consider a surface in $\mathbb{P}^3$ over $\mathbb{F}_q$ of the form 
\[ X: w^{r+1}=f_{r+1}(x, y, z), \]
where $f$ is a homogeneous polynomial in $x, y,$ and $z$ of degree $r+1$. The projection map sending $[x, y, z, w]$ to $[x, y, z]$ restricts to the morphism $\mathcal{X}$. Consider the curve $$C: f_{r+1}(x, y, z)=0,$$ called the branch locus. 
It can be guaranteed that 
 the distinguished fibers of $\mathcal{X}$ from points outside of $C$ have $r+1$ points \cite[Section 4.7]{BHHMV}. The inputs of the evaluation code are $\mathbb{F}_q$-points from 
\[ (X\setminus \varphi^{-1}(C))\cap U,\] where $U=\{p\in \mathbb{P}^3: p=[x, y, z, w], z\neq 0\}$. 

The examples discussed in \cite{BHHMV} include codes from cubic surfaces, where $f_3(x, y, z)$ is a homogeneous cubic polynomial, and K3 surfaces, where $f_4(x, y, z)$ is a homogeneous polynomial of degree four.  The authors also consider surfaces of general type, in which $f_5(x, y, z)$ is a homogeneous polynomial of degree five. Notable examples of code parameters are as follows. 

\begin{thm}{(Barg, Haymaker, Howe, Matthews, \& V\'{a}rilly-Alvarado 2017)}
There exist optimal LRCs designed from surfaces with lengths $q^2+2$, $q^2-1$, and $q^2-11$.   
\end{thm}

\begin{table}[h!]
\centering 
\begin{tabular}{|l|l|l|l|l|l|l|}
\hline
$q$ & $n$ & $k$ & $d$ & $r$ & Surface    & Reference \\ \hline
4   & 18  & 11  & 3   & 2   & Cubic      & Ex. 4.23  \\
5   & 24  & 17  & 3   & 3   & Quartic K3 & Ex. 4.26  \\
7   & 48  & 31  & 3   & 2   & Cubic      & Ex. 4.25  \\
11  & 110 & 87  & 3   & 4   & Quintic    & Ex. 4.27  \\ \hline
\end{tabular} \\ \ \\
\caption{Parameters of optimal codes from surfaces presented in \cite{BHHMV}.}
\label{table:lrc-surface}
\end{table}

The parameters of some optimal codes presented in \cite{BHHMV} are summarized in Table~\ref{table:lrc-surface}.


\section{Conclusion} 
\label{section:conclusion}
This article surveyed recent developments using underlying algebraic and geometric structures to create codes for local erasure recovery. However, some excellent work was not mentioned here due to space limitations. One of the strengths of these constructions is their design flexibility. Their generality makes them especially useful to accommodate changing needs.

\nocite{fiber, TB, TBV, lifted, GXY, BCGLP, SVV, 7922614}

\bibliography{glm_bib}{}
\bibliographystyle{abbrv}

\section*{Short Bios}

\noindent
Kathryn Haymaker is an Associate Professor of Mathematics in the Department of Mathematics and Statistics at Villanova University. She earned a Ph.D. in Mathematics from the University of Nebraska-Lincoln and a B.A. in Mathematics from Bryn Mawr College. Her research interests include coding theory and applications of discrete mathematics to problems in communications.\\

\noindent
Gretchen Matthews is a Professor of Mathematics at Virginia Tech and Director of a regional component of the Commonwealth Cyber Initiative (CCI). Matthews earned a B.S. from Oklahoma State University, a Ph.D. from Louisiana State University (both in mathematics), and an M.B.A. from Virginia Tech. She held a postdoctoral appointment at the University of Tennessee and was on the faculty at Clemson University. Her research interests include algebraic geometry and combinatorics and their applications to coding theory and cryptography.\\

\noindent
Hiram H. L\'opez is an Assistant Professor in the Department of Mathematics at Virginia Tech. He held positions as an Assistant Professor at Cleveland State University and as a Postdoctoral Fellow at Clemson University. He received a Ph.D. in mathematics from CINVESTAV-IPN and a B.S. in applied mathematics from the Autonomous University of Aguascalientes. His research interests include coding theory, commutative algebra, and image processing.\\

\noindent
Beth Malmskog is an Associate Professor of Mathematics at Colorado College. She held appointments as an Assistant Professor at Villanova University and Van Vleck Visiting Assistant Professor at Wesleyan University. She earned a BS in mathematics from the University of Wyoming and a Ph.D. in mathematics from Colorado State University.  Her research is in arithmetic geometry, number theory, coding theory, combinatorics, and mathematical aspects of fairness. \\

\noindent
Fernando Pi\~nero is an Assistant Professor at the University of Puerto Rico - Ponce. He received a B.Sc. and M.Sc. from the Unversity of Puerto Rico - Rio Piedras and a Ph.D. from the Technical University of Denmark, all in mathematics. He was a postdoctoral fellow at the Indian Institute of Technology - Bombay. His research interests include codes from algebraic geometry, Tanner codes, generalized LDPC codes, and linear codes from algebraic graphs.

\end{document}